\documentclass[a4paper,11pt]{article}
\usepackage{pos}
\usepackage{subcaption}
\usepackage{setspace}

\title{Study of a pulsar wind nebula candidate around the intermediate-age pulsar PSR J1413-6205 with H.E.S.S.}
\ShortTitle{Study of a PWN candidate around PSR J1413-6205 with H.E.S.S.}

\manuallySeparateAuthors
\author*[a]{Pauline Chambery}
\author[b]{, Yves Gallant}
\author[c,d,e]{, Armelle Jardin-Blicq}
\author[a]{, Marianne Lemoine-Goumard}
\author[c]{, Vincent Marandon}
\author[b,f]{, Atreyee Sinha}
\author[c,g]{, and Michelle Tsirou}
\author{ for the H.E.S.S. Collaboration}

\affiliation[a]{Université Bordeaux, CNRS/IN2P3, LP2I Bordeaux, UMR 5797,\\ F-33170 Gradignan, France}

\affiliation[b]{LUPM, Université de Montpellier, CNRS/IN2P3, CC 72,\\
Place Eugène Bataillon, F-34095 Montpellier Cedex 5, France}

\affiliation[c]{Max-Planck-Institut für Kernphysik,\\
P.O Box 103980, D-69029 Heidelberg, Germany}

\affiliation[d]{Department of Physics, Faculty of Science, Chulalongkorn University,\\
254 Phayathai Road, Pathumwan, Bangkok 10330, Thailand}

\affiliation[e]{National Astronomical Research Institute of Thailand (Public Organization)\\
Don Kaeo, MaeRim, Chiang Mai 50180, Thailand}

\affiliation[f]{Dept de EMFTEL, Instituto de IPARCOS, Universidad Complutense de Madrid\\
Madrid, Spain}

\affiliation[g]{Deutsches Elektronen-Synchrotron,\\
DESY, Zeuthen}

\emailAdd{chambery@cenbg.in2p3.fr}

\abstract{Very-high-energy $\gamma$-ray emission provides constraints on the morphology and the physics mechanisms involved in the evolution of pulsar wind nebulae (PWNe). In the Galactic plane, around $312 ^{\circ}$ of Galactic longitude, a promising region two-degree wide containing five powerful pulsars may offer a new insight on the transition between TeV-emitting PWNe and pulsar halos. Their rotational energies range from $10^{35}$ to $10^{37}$ erg s$^{-1}$ for ages between 13.6 and 62.8 kyr. Extended emission is detected with H.E.S.S. (High Energy Stereoscopic System) in their vicinity, notably around the pulsar PSR J1413-6205.

We processed 124 hours of H.E.S.S observations with an algorithm improving background fitting for the study of extended sources. We applied a three-dimensional likelihood analysis technique to model the different sources in the region using a configuration that optimizes the collection area at the highest energies.

This contribution focuses on the detection of a new extended source around PSR J1413-6205 over 5$\sigma$ with a hard spectrum. Preliminary results on this source show a radius of $0.12 ^{\circ}$ $\pm$ $0.01 ^{\circ}_{\rm stat}$, an index of 2.06 $\pm$ 0.20$_{\rm stat}$ and a lower limit on a cut-off energy of 17 TeV, at a 90\% confidence level. The detected emission is consistent with previous PWN models.}

\FullConference{%
  7th Heidelberg International Symposium on High-Energy Gamma-Ray Astronomy (Gamma2022)\\
  4-8 July 2022\\
  Barcelona, Spain\\}

\begin{document}
\maketitle

\setlength{\parindent}{0.8cm}

\section{Introduction}

The wide field of view and high sensitivity particle detector HAWC (High Altitude Water Cherenkov) reported extended emission at scales of a few degrees around the Geminga and Monogem pulsars \cite{HAWC_discovery}. This extended emission is suggestive of a "pulsar halo", interpreted as an evolutionary stage in which particles have escaped from the pulsar wind nebula (PWN) \cite{TeV_halo}. The very extended $\gamma$-ray emission from these halos is produced via escaping electrons and positrons scattered via inverse Compton on ambient interstellar radiation fields. In these systems, the pulsar is "middle-aged"\footnote{Throughout the paper we define middle-aged pulsars to be those with characteristic ages between 100-400 kyr \cite{TeV_halo}.} with a larger characteristic age and lower spin-down luminosity than the ones powering firmly identified PWNe in the TeV range \cite{PWN_pop_HGPS}. Thus, the pulsar is no longer a main contributor to the dynamics of the emission, as diffusion becomes one of the dominant processes \cite{halo_evolution}.

The analysis goal is to study PWN candidates likely to be transitioning to a "halo" in order to better understand their evolution and their impact on the Galactic cosmic-ray propagation. The Cherenkov telescopes of H.E.S.S., with a better angular and energy resolution compared to HAWC \cite{HAWC_HESS_perfs}, may allow to disentangle $\gamma$-ray sources in this region, while conducting spectro-morphological analyses of these extended-emission systems.

The region chosen for the analysis is centered at Galactic coordinates ($312 ^{\circ}$, $0 ^{\circ}$) with a diameter of two degrees. In this part of the sky, there are five pulsars with high spin-down luminosities and different ages, listed in Table \ref{5pulsars}, three of which are older than any firmly identified TeV-emitting PWN. We define these as intermediate-age\footnote{Between known TeV PWNe and pulsar halos} pulsars. Therefore, this region could contain observable PWNe at different stages of evolution, even up to the pulsar halo stage. In the H.E.S.S. Galactic Plane Survey (HGPS) \cite{HGPS}, no source was identified nor significantly detected in this region. New sky maps of this region have been made with twice the exposure time since the HGPS, showing significant extended emissions above 1 TeV, concentrated around PSR J1413-6205 beyond 5 TeV. We performed a more sophisticated analysis of the region, on the full dataset from 2004 to 2019, focusing on the emission around PSR J1413-6205, to understand its nature.

\noindent
\begin{center}
    \captionof{table}{Characteristic ages $\tau_{c}$ and spin-down powers $\dot{E}$ from the ATNF catalog v1.68 \cite{ATNF}.}
    \begin{tabular}{ccc}
        \hline
        PSR & $\tau_{c}$ (kyr)& $\dot{E}$ (erg s$^{-1}$) \\ \hline
        J1406-6121 & $61.7$ & $2.2 \times 10^{35}$ \\ \hline
        J1410-6132 & $24.8$ & $1 \times 10^{37}$ \\ \hline
        J1412-6145 & $50.4$ & $1.2 \times 10^{35}$ \\ \hline
        J1413-6141 & $13.6$ & $5.6 \times 10^{35}$ \\ \hline
        J1413-6205 & $62.8$ & $8.3 \times 10^{35}$ \\
    \hline
    \label{5pulsars}
    \end{tabular}
\end{center}

\setlength{\parindent}{0.8cm}

\section{Analysis method}

\subsection{Data selection and preparation}

Data were processed with the H.E.S.S. analysis package (HAP), applying a Hillas-type shower reconstruction \cite{Hillas} and the multi-variate analysis (MVA) technique \cite{MVA} for efficient $\gamma$-ray/hadron discrimination. Pre-selection and MVA discrimination cuts were optimized to improve the collection area at high energies (E > 1 TeV). The high-level analysis results were obtained using gammapy \cite{gammapy_1} \cite{gammapy_2}, a community-developed Python package for TeV $\gamma$-ray astronomy. All results presented here use twice the amount of data compared to the HGPS, in a radius of $4 ^{\circ}$ around the position of PSR J1413-6205, taken by the four 12-m telescopes with a zenith angle z < $60 ^{\circ }$. Data in the energy range 0.8-80 TeV were reduced by checking the run quality. In addition, events falling further than $2.3 ^{\circ}$ from the camera centre, or with energies lower than the abscissa of the background spectrum maximum were rejected. This selection removed the lowest energy bin, raising the analysis threshold to 1.3 TeV.

The results were independently cross-checked with data reduced using the Image Pixel-wise fit for Atmospheric Cherenkov Telescopes method \cite{ImPACT}, favoring the low-energy events collection.

\subsection{Analysis techniques}
\label{Analysis_techniques}

While analysing extended sources with possible energy-dependent morphologies, the challenge is to estimate the residual hadronic and the Galactic diffuse backgrounds. Consequently, the choice of exclusion masks for background parameters fitting is very important. We mask the known TeV sources \cite{gammacat} (HESS J1356-645, HESS J1418-609, HESS J1420-607, HESS J1427-608, RCW 86) and two additional regions. One with a radius of $0.4 ^{\circ}$ covers the surrounding of PSR J1413-6205 and one of $0.6 ^{\circ}$ the vicinity of the other four ATNF pulsars listed in Table \ref{5pulsars}. We checked that the significance distribution outside the exclusion regions follows a Gaussian, as predicted by Wilks' theorem \cite{Wilks}, with mean 0.07 and a sigma of 1.07.

To study such a complex region, a three-dimension (3D) spectro-morphological analysis \cite{gammapy_1} \cite{gammapy_2}, enabled by the gammapy tool, is paramount \cite{3D_analysis}. It allows us to model simultaneously the spatial and spectral properties of the sources in the region. In the source model definition, we freed all the parameters describing their morphology and spectrum. However, the positions were fixed to those of the pulsars, corresponding to excesses on the sky map, since releasing them did not significantly improve the fit. At the same time, a power-law spectral model for the background model of all stacked runs, described in the first paragraph of this subsection, is fitted. Further source components were added in a step-wise process. A component was added at the location of a visible significant excess on the sky map, as can be seen in Figure \ref{J1413_maps}. The component was kept in the model if it improved the fit by more than 5$\sigma$. Besides, different spatial and spectral models were tested for a component and kept if they improved the fit by at least 3$\sigma$.

\section{Results of extended $\gamma$-ray emission around PSR J1413-6205}

\subsection{Detection in a complex region}

The significance map, Figure \ref{J1413_maps}, shows the multiplicity of sources present in this region. The best 3D model describes this region, above 1 TeV, with several extended components. One of them extends around PSR J1413-6205 with a significance of more than 5$\sigma$ compared to the model without this component, for three degrees of freedom. This source is represented by a disk and a power-law spectrum. To understand the influence of the Galactic diffuse emission on the detection of the source coincident with PSR J1413-6205, we added a diffuse emission component for which we tried several models: the Fermi Galactic diffuse model \cite{4FGL_DR3}, a Galactic model candidate for CTA \cite{CTA_GPS} and a CO sky map model \cite{CO_model}. The addition of any of these models did not affect the parameters of this source, and even improved the likelihood of the fit. This indicates that the remaining emission in the region is a combination of unresolved sources and diffuse emission. A more detailed analysis characterizing these sources will be published soon.

\noindent\begin{minipage}[c]{0.47\textwidth} 
    \centering
    \includegraphics[width=\textwidth]{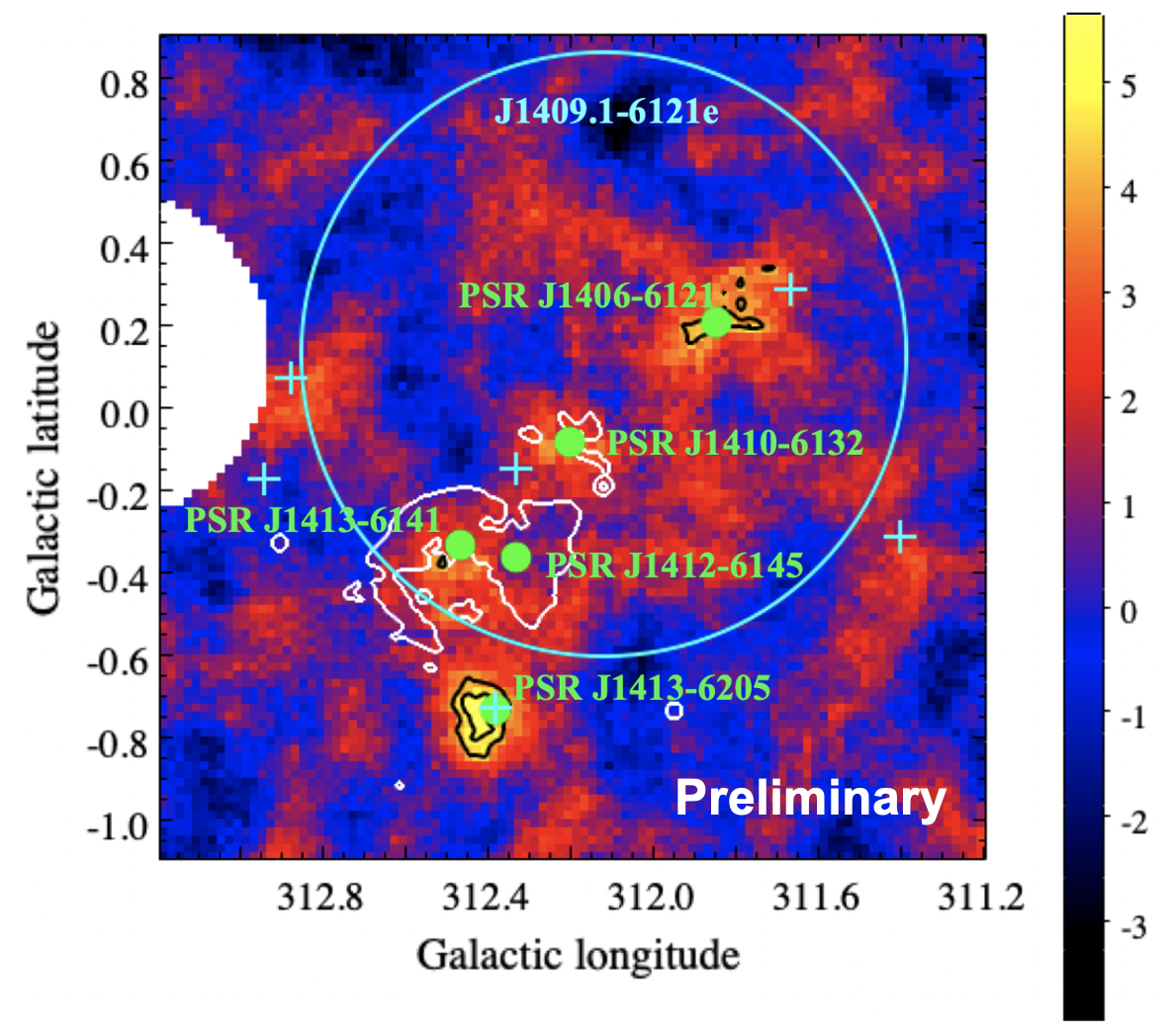}
\end{minipage}
\hfill
\begin{minipage}[c]{0.49\textwidth} 
    \centering
    \includegraphics[width=\textwidth]{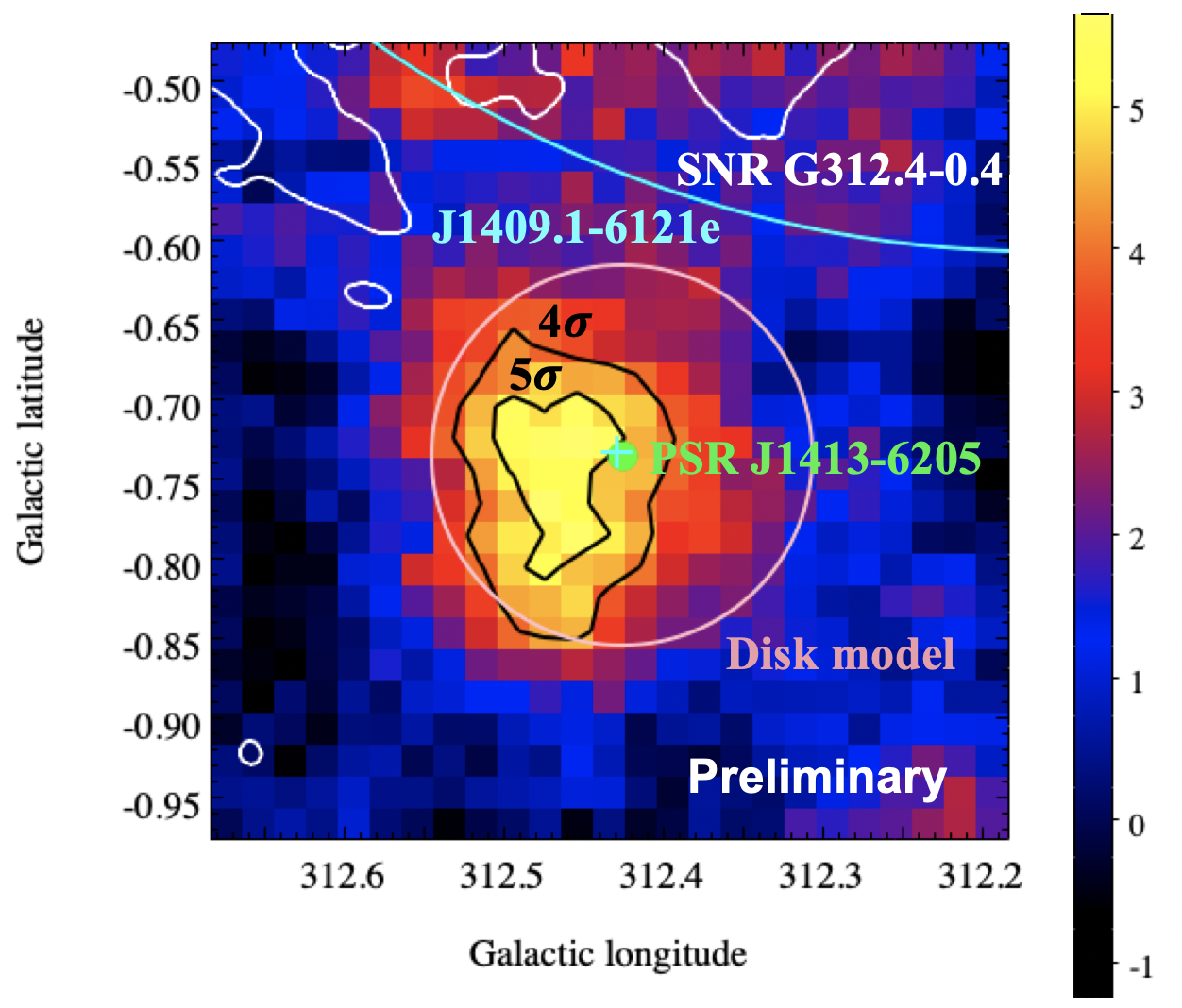}
\end{minipage}

\captionof{figure}{\label{J1413_maps} \emph{Left}: $2 ^{\circ} \times 2 ^{\circ}$ H.E.S.S. significance map ($0.1 ^{\circ}$ top-hat correlation), centred at ($312 ^{\circ}$, $0 ^{\circ}$) above 1.3 TeV. A white mask hides Kookaburra sources (HESS J1420-607 and HESS J1418-609). Turquoise crosses and circle corresponds to the sources from the 4FGL DR3 catalog \cite{4FGL_DR3}; white contours to the SNR G312.4-0.4 radio contours \cite{SNR_radio}; black contours to the H.E.S.S. 4$\sigma$ and 5$\sigma$ significance levels; and green dots to the pulsar positions from the ATNF catalog v1.68 \cite{ATNF}. \emph{Right}: Same map zoomed on $0.5 ^{\circ} \times 0.5 ^{\circ}$, centred on PSR J1413-6205. The pink circle is the best-fit extent of the disk, surrounding PSR J1413-6205.}

\setlength{\parindent}{0.8cm}

Figure \ref{J1413_maps} shows that the emission around PSR J1413-6205 is significant and the best-fit 3D analysis results confirm a significant extension. The preliminary radius disk found is $0.12 ^{\circ}$ $\pm$ $0.01 ^{\circ}_{\rm stat}$ around the pulsar position. Finally, assuming a PWN scenario, we looked at the energy dependence of the extension by refitting in smaller energy bands and its evolution was not significant.

\subsection{Spectral study of the source, a pulsar wind nebula behavior?}

\noindent\begin{minipage}[c]{0.5\textwidth}
    \centering
    \includegraphics[width=\textwidth]{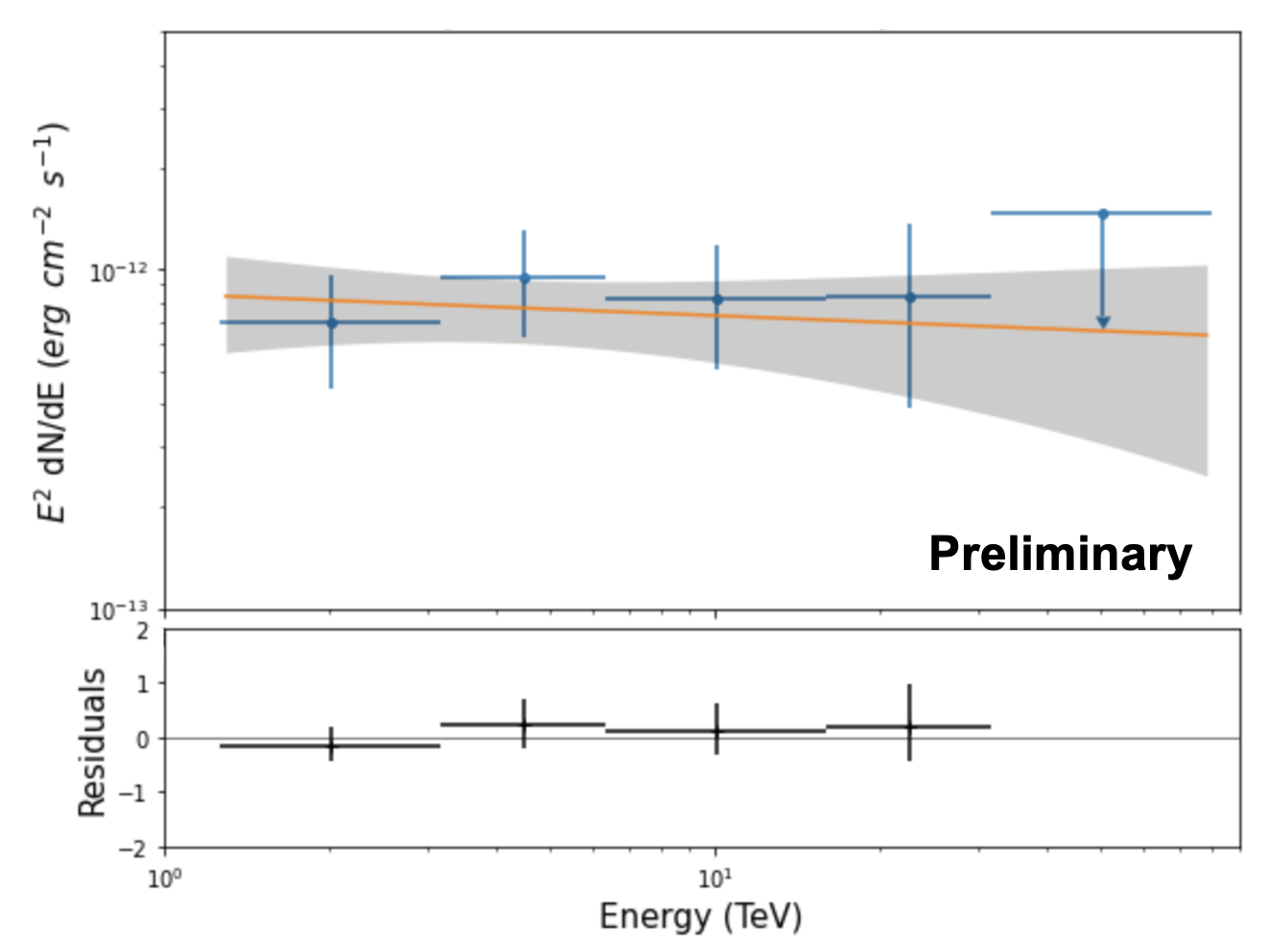}
\end{minipage}
\hfill
\begin{minipage}[c]{0.5\textwidth}
    \centering
    \begin{tabular}{|l|l|}
      \hline
      \multicolumn{2}{|c|}{Preliminary spatial parameters} \\
      \hline
        Pulsar position (l$^{\circ}$, b$^{\circ}$) & (312.37, -0.74) \\
        Disk radius ($^{\circ}$) & 0.12 $\pm$ 0.01$_{\rm stat}$ \\
        \hline
      \multicolumn{2}{|c|}{Preliminary spectral parameters} \\
      \hline
        Index & 2.06 $\pm$ 0.20$_{\rm stat}$ \\
        Norm (TeV$^{-1}$ cm$^{-2}$ s$^{-1}$) & (2.47 $\pm$ 0.52$_{\rm stat}$) $10^{-14}$ \\
        Pivot Energy (TeV) & 4.41 \\
        \hline
    \end{tabular}
\end{minipage}
\captionof{figure}{\label{J1413_SED_params} \emph{Left}: Spectral energy distribution of the extended source around PSR J1413-6205. \emph{Right}: Parameters of the 3D best model of this source, all parameters were left free in the fits except the position.}

\setlength{\parindent}{0.8cm}

The spectral energy distribution shown in Figure \ref{J1413_SED_params}, corresponds to the detected emission around PSR J1413-6205 from 1 TeV to tens of TeV, described by a power-law spectrum with a hard photon index. The preliminary spectral parameters of the source are an index of 2.06 $\pm$ 0.20$_{\rm stat}$, with a flux of $2.47\times10^{-14}$ TeV$^{-1}$ cm$^{-2}$ s$^{-1}$ at a pivot energy of 4.41 TeV. To further understand the nature of this source, we assumed an exponential cut-off power law. The lower limit on the cut-off energy of this source has been estimated at 90\% C.L. at 17 TeV. The hard index and lack of curvature of this source could indicate efficient acceleration.

\begin{figure}[!h]
    \centering
    \includegraphics[width=6.9cm]{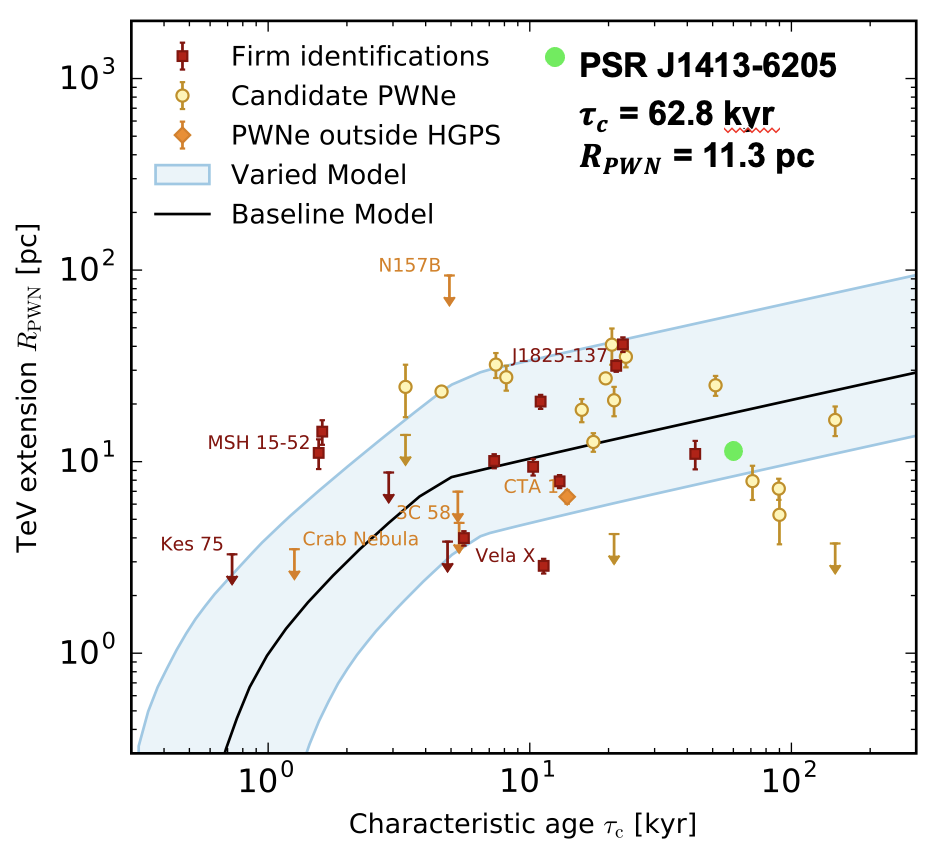}
    \includegraphics[width=7.1cm]{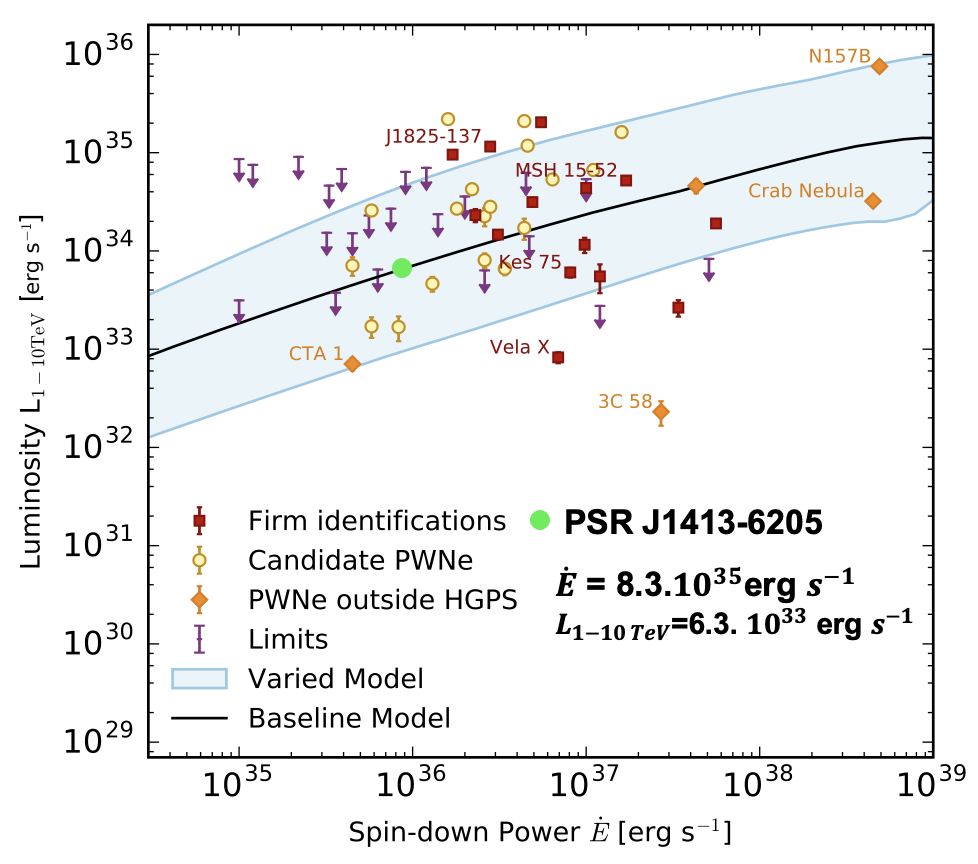}
    \caption{Plots adapted from the PWN population study in the HGPS \cite{PWN_pop_HGPS}. The green dot is associated with the extended emission around PSR J1413-6205.}
\label{PWN_pop}
\end{figure}

We considered the PWN population study performed with the HGPS \cite{PWN_pop_HGPS} (see Figure \ref{PWN_pop}) to assess the nature of our source of interest. The first plot was obtained with the pulsar characteristic age $\tau_{c}$ from the ATNF catalog v1.68 \cite{ATNF}, and the TeV extension $R_{PWN}$, calculated from the 3D disk model source radius and the pulsar distance. In the absence of a robust radio distance measurement, we based our choice on a strong hypothesis which is that the pulsar at longitude $312 ^{\circ}$ is close to the Galactic spiral arm tangent distance \cite{galactic_arm}. For a solar Galactocentric radius of 8.0 kpc, this distance is approximately 5.4 kpc. The second plot was obtained with the pulsar spin-down power $\dot{E}$ from the ATNF catalog v1.68 \cite{ATNF}, and the luminosity $L_{1-10\mbox{\tiny TeV}}$ which was calculated from the 3D-analysis integrated photon flux above 1 TeV and the pulsar distance. The study shows that the analysed source follows the trend of PWN models. Nevertheless, the age of PSR J1413-6205 approaches "middle-aged" pulsars, no energy-dependent morphology has been detected for the source around this pulsar and we do not yet have strict enough criteria to fully discriminate a PWN from a pulsar halo. This source could also be in the transition stage to a pulsar halo.

\section{Conclusion}

We detect significant extended emission close to the intermediate-age pulsar PSR J1413-6205 with a hard spectrum. This source could be an efficient particle accelerator and is fully consistent with current PWN models. More statistics would allow us to see if there is a variation of its morphology with energy to determine its evolutionary stage with respect to the halo phase, and constrain the maximum energy of the accelerated particles. In addition, a joint analysis of Fermi-LAT and H.E.S.S. data would allow us to better understand the nature of this source, to determine whether protons or electrons produce the $\gamma$-ray emission and to confirm the detection of several other new sources in this field of view. Systematic errors will thus be considered for the analysis with both instruments at the same time. With the improved performances of CTA (Cherenkov Telescope Array) \cite{CTA_performances}, we will be able, in such regions, to discriminate the extended sources and to discover objects at different phases from PWN to pulsar halos \cite{halos_everywhere}.\newline

\footnotesize
\noindent\emph{Acknowledgements}. PC and MLG acknowledge support by ANR for the GAMALO project under reference ANR-19-CE31-0014, and for the PECORA project under reference ANR-17-CE31-0014. AS acknowledges funding from Spain´s Minsitry of Universities through the Maria Zambrano Talent Attraction Programme 2021-2023, financed by NextGenerationEU.

\begin{spacing}{0.8}

\end{spacing}


\begin{thebibliography}{99}

\bibitem{HAWC_discovery} A.U. Abeysekara, A. Albert, R. Alfaro et al., \emph{Extended $\gamma$-ray sources around pulsars constrain the origin of the positron flux at Earth}, Science, 358, 911, 2017.

\bibitem{TeV_halo} T. Linden et al., \emph{Using HAWC to Discover Invisible Pulsars}, Phys. Rev. D 96, 103016, 2017.

\bibitem{PWN_pop_HGPS} H.E.S.S. Collaboration, \emph{The population of TeV PWNe in the HGPS}, A\&A 612, A2, 2018.

\bibitem{halo_evolution} G. Giacinti et al., \emph{Halo fraction in TeV-bright pulsar wind nebulae}, A\&A 636, A113, 2020.

\bibitem{HAWC_HESS_perfs} H. Abdalla, F. Aharonian, F. Ait Benkhali, et al., \emph{TeV Emission of Galactic Plane Sources with HAWC and H.E.S.S.}, ApJ, 917, 6, 2021.

\bibitem{HGPS} H.E.S.S. Collaboration, \emph{The H.E.S.S. Galactic plane survey}, A\&A 612, A1, 2018.

\bibitem{ATNF} Manchester R. N., Hobbs G. B., Teoh A., Hobbs M., \emph{The Australia Telescope National Facility Pulsar Catalogue}, The AJ 129, Issue 4, pp. 1993-2006, 2005. URL: http://www.atnf.csiro.au/research/pulsar/psrcat.

\bibitem{Hillas} A. M. Hillas, \emph{Cerenkov Light Images of EAS Produced by Primary Gamma Rays and by Nuclei}, ICRC1985, Vol 3, p.445, 1985.

\bibitem{MVA} Y. Becherini et al., \emph{Advanced analysis and event reconstruction for the CTA Observatory}, Heidelberg Symposium on High Energy Gamma-Ray Astronomy, 2012.

\bibitem{gammapy_1} C. Deil and R. Zanin et al., \emph{Gammapy - A prototype for the CTA science tools}, ICRC2017, 301, 766, 2017.

\bibitem{gammapy_2} A. Donath et al., \emph{gammapy/gammapy: v.0.19}, Zenodo, 2021. DOI: 10.5281/zenodo.5721467. URL: https://doi.org/10.5281/zenodo.5721467.

\bibitem{ImPACT} R.D. Parsons and J.A. Hinton, \emph{A Monte Carlo Template based analysis for Air-Cherenkov Arrays}, Astroparticle Physics, 56, 26-34, 2014.

\bibitem{3D_analysis} L. Mohrmann et al., \emph{Towards a 3D likelihood analysis in very-high-energy $\gamma$-ray astronomy: the case of H.E.S.S.}, PoS, ICRC2019, 747, 2019.

\bibitem{gammacat} A. Voruganti, A. Donath, C. Deil, G. Maier, M. Wegen and P. Deiml, 2018. \emph{gamma-cat} URL: https://gamma-cat.readthedocs.io/.

\bibitem{Wilks} S. S. Wilks, \emph{The large-sample distribution of the likelihood ratio for testing composite hypotheses}, The Annals of Mathematical Statistics, 9, 60–62, 1938.

\bibitem{4FGL_DR3} Fermi-LAT collaboration, \emph{Incremental Fermi Large Area Telescope Fourth Source Catalog}, ApJS 260, 53, 2022.

\bibitem{CTA_GPS} Q. Remy, \emph{Survey of the Galactic Plane with the CTA}, PoS, ICRC2021, 2021.

\bibitem{CO_model} T. M. Dame, D. Hartmann, P. Thaddeus, \emph{The Milky Way in Molecular Clouds: A New Complete CO Survey}, The AJ, 547, 792-813, 2001.

\bibitem{SNR_radio}  J. L. Caswell and P. J. Barnes, \emph{A new galactic supernova remnant, G312.4–0.4}, Mon. Not. R. astr. Soc., 216, 3, 753–760, 1985.

\bibitem{galactic_arm} J. P. Vallée, \emph{A New Multi-Tracer Approach to Defining the Spiral arm width in the Milky Way}, The AJ, 896, 19, 1-10, 2020.

\bibitem{CTA_performances} CTA collaboration, CTAO Performance, 2016. URL: https://www.cta-observatory.org/science/ctao-performance/.

\bibitem{halos_everywhere} Sudoh et al., \emph{TeV Halos are Everywhere: Prospects for New Discoveries}, Phys. Rev. D 100, 043016, 2019.

\end{thebibliography}
\end{document}